\documentclass[aps,pra,
amsmath,
amssymb,showpacs,showkeys,reprint]{revtex4-1}

\usepackage{graphicx}
\usepackage{bm}
\usepackage{epstopdf}
\usepackage{float}

\begin{document}
\preprint{APS/123-QED}

\title{Principles of an Atomtronic Battery}

\author{Alex A. Zozulya}\email[]{zozulya@wpi.edu}
\affiliation{Department of Physics, Worcester Polytechnic Institute,
100 Institute Road, Worcester, Massachusetts 01609, USA}

\author{Dana Z. Anderson}
\affiliation{Department of Physics and JILA, University of Colorado
and National Institute of Standards and Technology, Boulder,
Colorado 80309-0440, USA}

\date{\today}

\begin{abstract}
An asymmetric atom trap is investigated as a means to implement a \textquotedblleft{battery}\textquotedblright that supplies ultracold atoms to an atomtronic circuit.  The battery model is derived from a scheme for continuous loading of a non-dissipative atom trap proposed by Roos et al.(Europhysics Letters $ \mathbf{61}$, 187 (2003)). The trap is defined by longitudinal and transverse trap frequencies $f_{z}$, $f_{\perp}$ and corresponding trap energy heights $U_{z}$, $U_{\perp}$. 
The battery's ability to supply power to a load is evaluated as a function of an input atom flux and power, $I_{in}$, $P_{in}=I_{in}(1+\epsilon)U_{z}$, where $\epsilon$ is an excess fractional energy.  For given trap parameters, the battery is shown to have a resonantly optimum value of $\epsilon$.  The battery behavior can be cast in terms of an equivalent circuit model;  specifically, for fixed input flux and power the battery is modeled in terms of a Th\'{e}venin equivalent chemical potential and internal resistance.  The internal resistance establishes the maximum power that can be supplied to a circuit, the heat that will be generated by the battery, and that noise will be imposed on the circuit.  We argue that \emph{any} means of implementing a battery for atomtronics can be represented by a Th\'{e}venin equivalent and that its performance will likewise be determined by an internal resistance.
\end{abstract}


\maketitle

\section{Introduction}
Atomtronics is an analog of electronics in which chemical potential and atom flux are the duals to electric voltage and current~\cite{Stickney:2007ix,Seaman:2007kx,Pepino:2009jb,Gajdacz:2012ux,Pepino:2010p5427}. Interest in atom-based devices and circuits is both academic and practical.  It is of fundamental interest, for example, to study ideal atom-based semiconductor material and device analogs that can be implemented using optical lattices.  On the practical side, atomtronics is of interest for ultracold atom based systems for inertial and magnetic field sensing, and more generally for quantum signal processing.  The analogy between electronics and atomtronics is sufficiently complete that it is reasonable to consider the development of atomtronic circuits that resemble electronic versions but operate with atoms and often in the quantum regime.  Circuits require a supply of energy to operate, of course.   Our objective in this paper is to elucidate some fundamental aspects of the atomtronic dual of an electronic power supply --call it simply a ``battery''.  We think of a battery as a self-contained component that supplies both power and particles to a load.  In our case, the particles are ultracold atoms.

From a physics perspective a battery can be assessed in terms of its ability to perform work. An idealized electrical battery maintains a voltage across its terminals whose value is independent of the load attached to its terminals.  A real battery, however, has an internal resistance that causes the voltage across the battery's terminals to drop when current is supplied to the load. Moreover, as the battery supplies current, power is also dissipated as heat in the internal resistance.  The notion of a battery as a component encapsulates its electrical function without regard for the electrochemistry that takes place ``under the hood," so to speak.  An electronic battery has some value of internal resistance no matter what the details of its electrochemistry; in fact, so does any supply of electrical power, chemically based or other.  Knowing enough about the underlying chemistry, physics, circuit design, etcetera, one could determine the internal resistance, at least in principle.

One should expect that, true to its electronic analog, an atomtronic battery will \textit{necessarily} have internal resistance.  And like the electronic case, the battery potential and internal resistance determine the maximum power $P_{max}$  that can be delivered to a load, the noise (the atomtronic equivalent of Johnson noise~\cite{Robinson:1974ww}) generated by fluctuating current in the internal resistance operating at finite temperature, as well as the heat dissipated by the internal resistance.  In the ultracold and quantum realm of atomtronics, one can appreciate that these aspects of the atomtronic battery and their impact on circuit behavior are of considerable interest.

In the following section we analyze a specific physical model for an atomtronic battery.  Keeping in mind that the battery's job is to supply power, our nominal intent is to characterize the dependence of chemical potential and the ability to provide ultracold atom flux to a load on the model's parameters.  As a circuit element we show that the battery can indeed be represented in terms of a Th\'{e}venin equivalent chemical potential and an internal series resistance, given a fixed set of model parameters.   While the analysis and results are specific to our model, the general conclusion is not: an element or sub circuit that supplies power to an atomtronic circuit will be  accompanied by an internal resistance that generates heat, introduces noise and limits the power available to the load regardless of the operational physics.
\section{Battery Model}\label{thermal_cloud}
\begin{figure}[h]
\includegraphics[width=8.6cm]{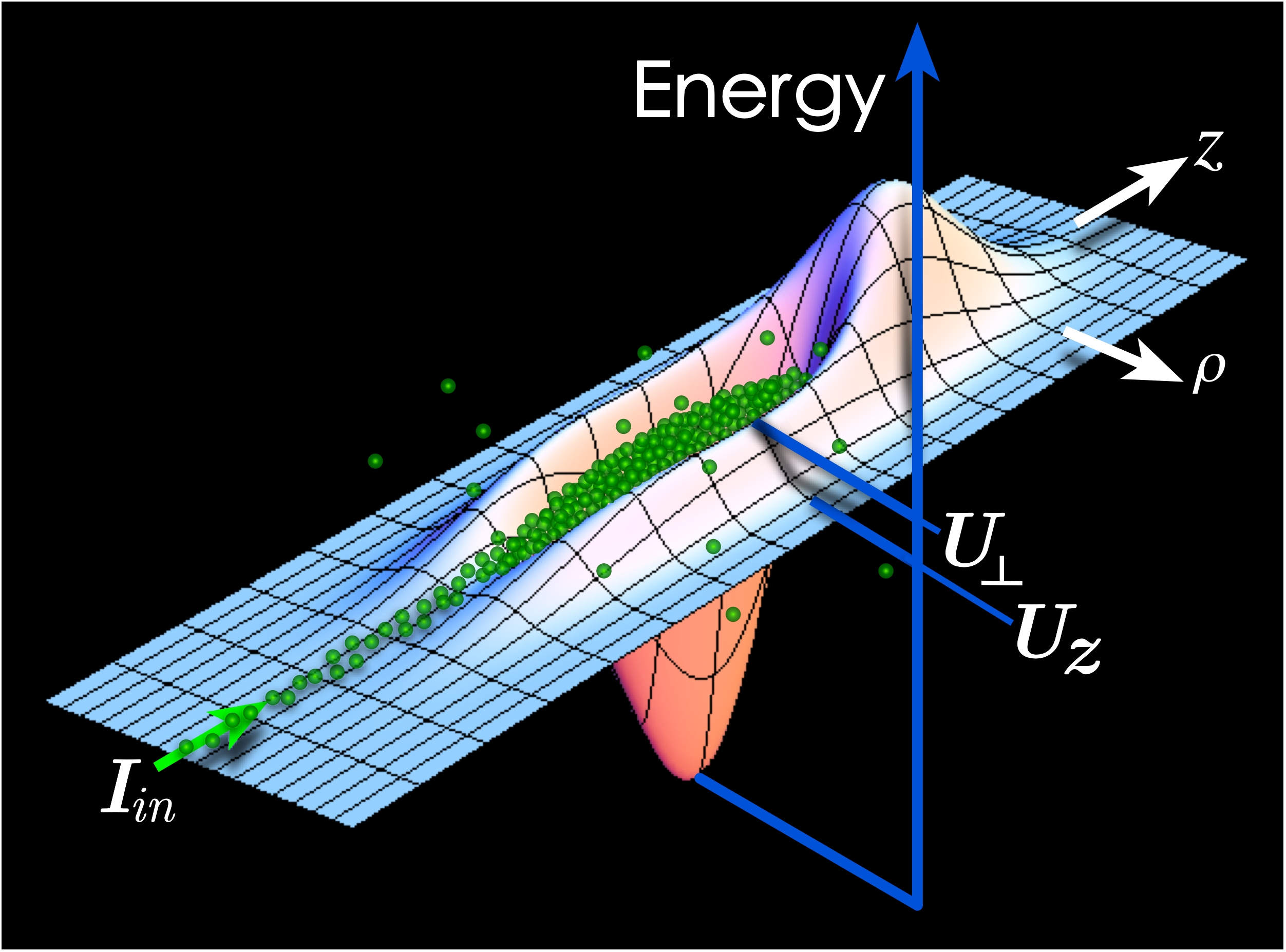}
\caption{\label{fig:potential} The battery potential is highly asymmetric, somewhat resembling a gravy bowl. Atoms enter along the longitudinal direction into the pour-spout of the bowl.  They scatter from the cloud of already trapped atoms, eventually reaching an equilibrium distribution and number as they also escape out the sides and input direction.}
\end{figure}

%
Our battery model is derived from a scheme for continuous loading of a non-dissipative atom trap proposed by Roos et al. in Ref.~\cite{Roos:2003tp}.  The work analyzes a highly asymmetric, cigar-shaped trapping potential subject to an incoming beam of cold atoms.  Fig.~\ref{fig:potential}  provides a conceptual illustration of the scheme.  Atoms enter the trapping region from the $-z$ (longitudinal) direction.  The height of the trapping potential along this direction is $U_{z}$ and the mean energy of the beam is $(1+\epsilon)U_{z}$.  A heightened potential at the far end prevents atoms from escaping out the +z direction.  Atoms from the beam are captured by undergoing collisions with the atoms already present in the trap.
Cooling of the trapped atoms is provided by their evaporation through the side walls and end cap of the trapping potential.
The heights of the trap along the transverse and longitudinal directions are equal to $U_{\perp}$ and $U_{z}$, respectively, with $U_{\perp} > U_{z}$.
The confinement frequency  $f_{\perp}$ along the two transverse dimensions $x$ and $y$ is much larger than that along the $z-$ axis:
$f_{\perp} \gg f_{z}$. Since the beam enters the trap along its longitudinal dimension which has low frequency, and, as a consequence, large size, it is completely absorbed in the trap due to collisions with the atoms inside.
The trap is populated by a cloud of $N_{ex}$ thermal atoms in contact with $N_{a}$ Bose-condensed atoms. Dynamics of non-condensed atoms in the trap is described by the set of equations (cf. \cite{Roos:2003tp})
\begin{equation}\label{dNexdt}
    \frac{d N_{ex}}{dt} =I_{in}- \left(p_{\perp} + p_{z}\right)\gamma N_{ex} - I_{l},
\end{equation}
\begin{eqnarray}\label{dEexdt}
{\frac{dE_{ex}}{dt}} =  &&I_{in}(1 + \epsilon)U_{z} - p_{z}(U_{z} + \kappa_{z}kT)\gamma N_{ex}\nonumber\\
  &&- p_{\perp}(U_{\perp} +  \kappa_{\perp}kT)\gamma N_{ex} - \mu_{a}I_{l}.
\end{eqnarray}
Here $N_{ex}$ and $E_{ex}$ are the number and energy of the thermal atoms in the trap, $I_{in}$ is the flux of incident atoms,
$\gamma$ is the average collision rate and $p_{z}$ and $p_{\perp}$ are the probabilities of evaporating after collision through the side and
end walls of the trap. The quantity $(1 + \epsilon)U_{z}$ is the average energy per incident atom, $U_{\perp}$ and $U_{z}$ are the evaporation thresholds for the perpendicular and longitudinal direction,
respectively, and $\kappa_{\perp} kT$ and $\kappa_{z}kT$ determine excess average energy per atom carried away during evaporation through the side or the z direction. Finally, $I_{l}$ is the rate of decrease or increase of the number of noncondensed atoms due to their interaction with the BEC. It is given by the expression \cite{Gardiner:97}
\begin{equation}\label{I_load}
    I_{l} = \frac{(8\pi)^{2}m(a_{s}kT)^{2}}{h^{3}}\frac{\mu_{ex} - \mu_{a}}{kT}N_{a},
\end{equation}
where $\mu_{a}$ and $\mu_{ex}$ are chemical potentials of the BEC and the thermal atoms in the well, $m$ is the atomic mass and $a_{s}$ the s-wave scattering length. Equation (\ref{I_load}) is written in the limit
$\mu_{a}, \mu_{ex} \ll kT$.
In the steady-state analysis carried out below $I_l$ is the current supplied to a load attached to the battery.  In other words, without specifying the details of how they do so, we presume that the elements attached to the battery extract only condensed atoms. Equation (\ref{I_load}) can be written in the form of an Ohm's law:
\begin{equation}\label{internalresistance}
I_l=\frac{\mu_{ex}-\mu_a}{R_a},
\end{equation}
where
\begin{equation}\label{R_a}
R_a= \frac{h^{3}}{(8\pi)^{2}m a_{s}^{2} N_{a}kT}.
\end{equation}

We shall see that $R_a$ contributes to the total internal resistance of the battery.

The BEC in the well is in Thomas-Fermi regime.  In case of a parabolic well, its chemical potential is given by the expression:
\begin{equation}\label{mu_a}
       \mu_{a} = \frac{15^{2/5}}{2}\left(\frac{N_{a}a_{s}}{\bar{a}}\right)^{2/5}h\bar{f},
\end{equation}
where $\bar{f} = (f_{\perp}^{2}f_{z})^{1/3}$ and $a = (h/4\pi^2m\bar{f})^{1/2}$.

The energy of the thermal atoms is given by the expression
\begin{eqnarray}
    E_{ex} \approx 3\frac{(kT)^{4}}{(h\bar{f})^{3}}\left[\zeta(4) + 3\frac{\mu_{ex}}{kT}\zeta(3)\right],
\end{eqnarray}
where $\mu_{ex}$ is the chemical potential of the thermal atoms determined by the expression
\begin{equation}\label{mu_ex}
    \frac{\mu_{ex}}{kT} \approx \frac{1}{\zeta(2)}\left[N_{ex} \left(\frac{h\bar{f}}{kT}\right)^{3} - \zeta(3)\right],
\end{equation}
and $\zeta$ is Riemann zeta-function.

Following Roos et. al \cite{Roos:2003tp}, we are assuming that the evaporation in the transverse direction takes place in the collisionless (Knudsen) regime $f_{\perp} \gg \gamma$, i.e.,
an atom emerging after a collision with an energy $E$ larger than the transverse evaporation energy $U_{\perp}$, leaves the trap
without undergoing any more collisions. The evaporation in the longitudinal direction takes place in the opposite limit $f_{z} \ll \gamma$.
This considerably reduces the evaporation rate along the $z-$ axis as compared with the rate derived in the collisionless regime for the same ratio $U/kT$.
Roos et al. discovered that the presence of the two evaporation channels results in a resonance in the steady state phase space density of the thermal atoms when the transverse
evaporation threshold $U_{\perp}$ is close to the incident energy of the atomic beam $(1 + \epsilon)U_{z}$.

We adopt the expressions for the probability of evaporation in the transverse and longitudinal directions derived in \cite{Roos:2003tp} from molecular-dynamics simulations.
The probability of evaporation in the transverse direction $p_{\perp}$ is given by the relation
\begin{equation}\label{p_perp_and_kappa_perp}
    p_{\perp} \simeq 2 e^{-\eta_{\perp}}, \; \kappa_{\perp} \simeq 2.0,
\end{equation}
where $\eta_{\perp} = U_{\perp}/kT$. Equation (\ref{p_perp_and_kappa_perp}) is valid in the range $8 < \eta_{\perp} < 13$.

The probability of evaporation in the longitudinal direction $p_{z}$ are given by the relation
\begin{equation}\label{p_z_and_kappa_z}
    p_{z} \simeq 0.14e^{-\eta_{z}}\frac{f_{z}}{\gamma}, \; \kappa_{z} \simeq 2.9,
\end{equation}
where $\eta_{z} = U_{z}/kT$. Equation (\ref{p_z_and_kappa_z})  assumes $f_{z} \ll \gamma$ and holds in the range $4 < \eta_{z} < 7$.

In the presence of the BEC the density of the thermal atoms above the BEC is fixed at the level
\begin{equation}\label{gen:thermal_density_inside_BEC}
    n_{ex}(\bm r) = \frac{1}{\Lambda^{3}}\zeta\left(\frac{3}{2}\right).
\end{equation}
where
\begin{equation}
    \Lambda = (h^{2}/2\pi mkT)^{1/2}
\end{equation}
is the thermal de Broglie wavelength.

The mean collision rate $\gamma$ in this case can be evaluated by the relation
\begin{equation}\label{mean_collision_rate_BEC}
    \gamma = {32\pi^2\zeta(3/2)}\frac{m(a_{s}kT)^{2}}{h^{3}}.
\end{equation}
%
\section{Steady state}
%
In the following we shall consider steady-state solutions of Eq.~(\ref{dNexdt}) and (\ref{dEexdt})

\begin{eqnarray}\label{temp1}
   I_{in}= &\left(p_{\perp} + p_{z}\right)\gamma N_{ex} + I_{l}, \nonumber \\
    I_{in}(1 + \epsilon)U_{z} =& p_{z}(U_{z} + \kappa_{z}kT)\gamma N_{ex} \nonumber \\
    +&p_{\perp}(U_{\perp} + \kappa_{\perp}kT)\gamma N_{ex}.
\end{eqnarray}
Solutions of Eqs.~(\ref{temp1}) will be analyzed for both zero and nonzero value of $I_{l}$, the last situation corresponding
to the case when the BEC is outcoupled from the trap at the rate $I_{l} = const.$ Since $U_{z} \gg \mu_{a}$, the term with $I_{l}$
in the second of Eq.~(\ref{temp1}) has been neglected.
The set of Eqs. (\ref{temp1}) can be transformed to the form
\begin{equation}\label{steady_state_N}
    N_{ex} = \frac{I_{in} - I_{l}}{\left(p_{\perp} + p_{z}\right)\gamma},
\end{equation}
\begin{equation}\label{steady_state_T}
    (1 + \epsilon^{\prime})U_{z} = \frac{p_{z}}{p_{\perp} + p_{z}}(U_{z} + \kappa_{z}kT) + \frac{p_{\perp}}{p_{\perp} + p_{z}}(U_{\perp} + \kappa_{\perp}kT),
\end{equation}
where
\begin{equation}\label{epsilon_prime}
    \epsilon^{\prime} = \frac{\epsilon + I_{l}/I_{in}}{1 - I_{l}/I_{in}}
\end{equation}
Equations (\ref{steady_state_N}) and (\ref{steady_state_T}) show that the nonzero rate of outcoupling $I_{l}$ is equivalent to
the situation with $I_{l} = 0$  but for a reduced flux of incident atoms $I_{in} - I_{l}$ and an increased mean energy of the incident beam. The changes are such that the input power is constant, i.e.,  $(I_{in} - I_{l})(1+\epsilon^{\prime})U_{z}=P_{in}{\equiv}I_{in}(1+\epsilon)U_{z}$.

The steady-state values of $T$ and $N_{ex}$ given by Eqs.~(\ref{steady_state_N}) and (\ref{steady_state_T}) depend on the ratio of the transverse and the
longitudinal evaporation energies $U_{\perp}/U_{z}$.

\begin{figure}[h]
\includegraphics[width=8.6cm]{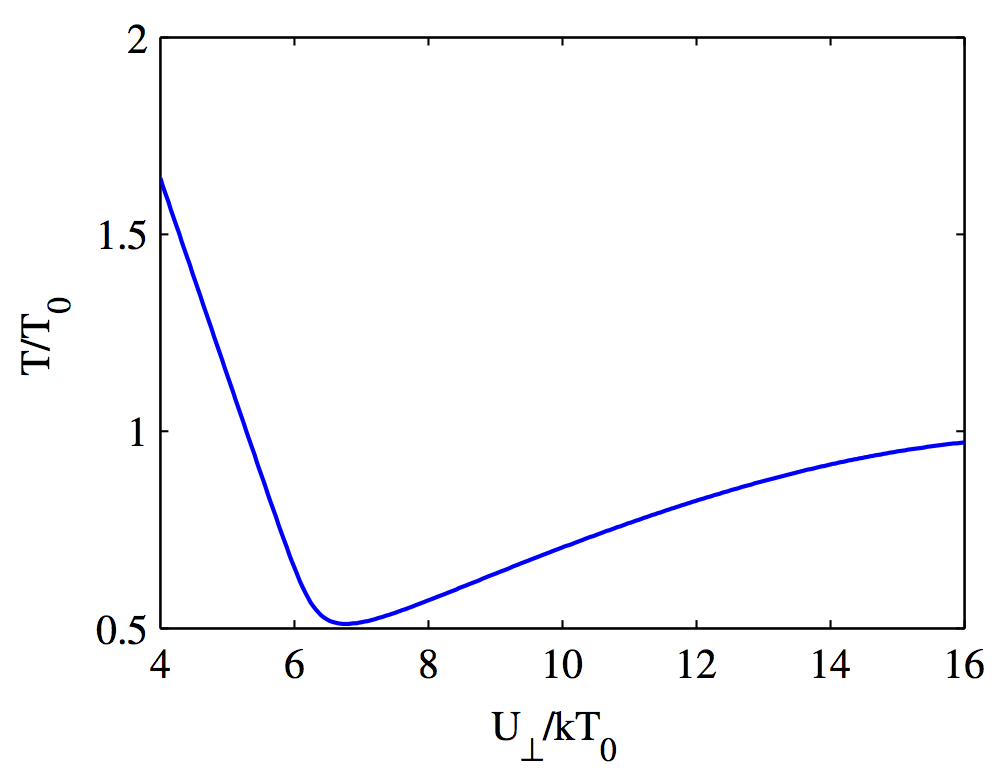}
\caption{\label{fig:T_versus_U_p} Normalized temperature $T/T_{0}$ versus transverse evaporation threshold $U_{\perp}/k T_{0}$}
\end{figure}

\begin{figure}[h]
\includegraphics[width=8.6cm]{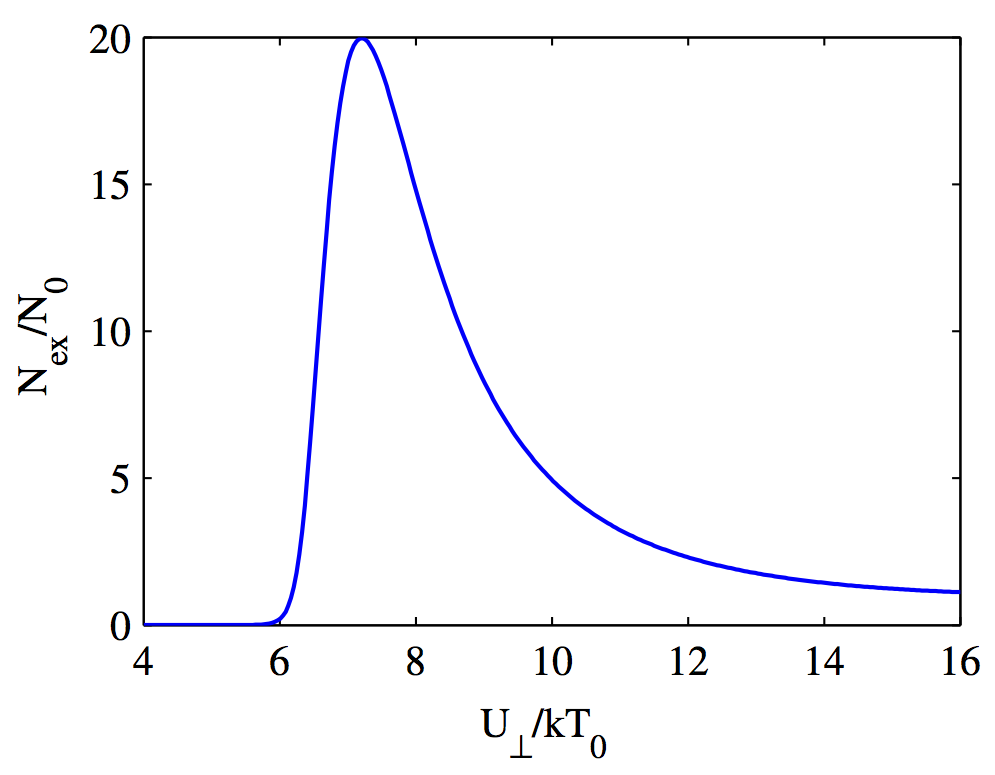}
\caption{\label{fig:N_versus_U_p} Normalized number of thermal atoms $N_{ex}/N_{0}$ versus transverse evaporation threshold $U_{\perp}/k T_{0}$}
\end{figure}

A convenient reference point is given by the values of the temperature $T_{0}$ and the population $N_{0}$ in the limit $U_{\perp} \gg U_{z}$ and for zero value of the load:
\begin{equation}\label{T0}
    kT_{0} = \frac{\epsilon U_{z}}{\kappa_{z}},
\end{equation}
\begin{equation}\label{N0_general}
    N_{0} = \frac{I_{in}}{p_{z}(T_{0})\gamma(T_{0})}.
\end{equation}
In particular, if $p_{z}$ is given by Eq.~(\ref{p_z_and_kappa_z}), Eq.~(\ref{N0_general}) can be written as

\begin{equation}\label{N0}
    N_{0} = 1.14 \frac{I_{in}}{f_{z}}\exp(\kappa_{z}/\epsilon).
\end{equation}
A typical dependence of $T$ and $N_{ex}$ on $U_{\perp}$ for all other parameters fixed is illustrated by Fig.~\ref{fig:T_versus_U_p} and Fig.~\ref{fig:N_versus_U_p}.
The number of thermal atoms and the temperature in Figs.~\ref{fig:T_versus_U_p} and \ref{fig:N_versus_U_p} are normalized to $T_{0}$ and $N_{0}$ given by the relations (\ref{T0}) and (\ref{N0}),
respectively. The parameters for Figs.~\ref{fig:T_versus_U_p} and \ref{fig:N_versus_U_p} are $\epsilon = 0.7$ and $\gamma_{0}/2{\pi}f_{z} = 100$. The optimum value of the transverse evaporation threshold
$U_{\perp}$ is equal to about $1.7U_{z}$ (the optimums for the temperature and the number of atoms are slightly different).
Relation (\ref{T0}) shows that the temperature of thermal atoms in the trap in the limit $U_{\perp} \gg U_{z}$ is pretty much
equal to to the excess energy $\epsilon U_{z}$ of the incident beam (the parameter $\kappa_{z}$ is of the order of one).
It seems that decreasing $\epsilon$ is the easiest way to dramatically lower the temperature of the thermal atoms in the trap,
but the situation is not so simple. Atoms in the incident beam have some characteristic thermal energy spread $kT_{\mathrm{i}}$.
This quantity should be of the order or smaller than the excess energy $\epsilon U_{z}$ of the incident beam.
In the opposite case $kT_{\mathrm{i}} > \epsilon U_{z}$, the excess energy of the beam entering the trap will be determined not by the $\epsilon U_{z}$, but by the
characteristic thermal width of the beam $kT_{i}$. The minimum temperature (optimized with respect to $U_{\perp}$) differs from its
asymptotic value (\ref{T0}) by a factor of two, so the temperature of the atoms in the trap can not be significantly lower than the
temperature of the incident atoms. For example, in the analysis of Ref.~\cite{Griffin:2009wv}, the temperature of the incident
atoms was $17~\mu\mathrm{K}$ and that of the atoms in the trap in the optimum regime was about $24\mu\mathrm{K}$.
\begin{figure}[h]
\includegraphics[width=8.6cm]{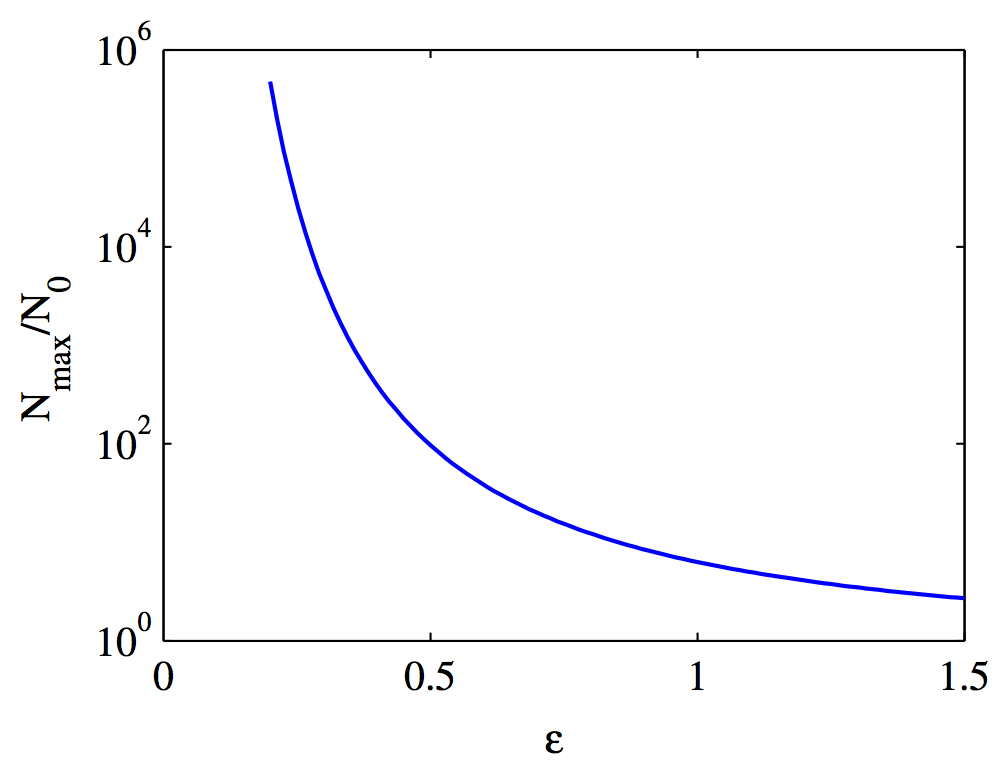}
\caption{\label{fig:N_max_versus_epsilon} Resonant enhancement factor $N_{max}/N_{0}$ versus $\epsilon$.}
\end{figure}
Changing $\epsilon$ has a much more dramatic effect on the number of trapped atoms under the optimum conditions.  Figure~\ref{fig:N_max_versus_epsilon} shows the resonant enhancement factor $N_{max}/N_{0}$ for different values of $\epsilon$. Here $N_{max}$ is the maximum number of atoms corresponding to the optimum choice
of $\epsilon$. For example, Fig.~\ref{fig:N_versus_U_p} shows a resonant increase in the number of atoms of about $100$ for $\epsilon = 0.5$. Increasing $\epsilon$ to $0.7$ decreases the ratio $N_{max}/N_{0}$ to $20$ and decreasing $\epsilon$ to $0.3$ increases $N_{max}/N_{0}$ to about $4 \times 10^{3}$.
This behavior is explained by Eq.~\ref{steady_state_N}:
\begin{eqnarray*}
    N_{ex} = \frac{I_{in}}{\left(p_{\perp} + p_{z}\right)\gamma}.
\end{eqnarray*}
The probabilities $p_{\perp}$ and $p_{z}$ depend on the temperature exponentially and even, say, a two-fold decrease in the asymptotic temperature $T_{0}$
(due to a two-fold decrease in $\epsilon$) strongly changes the number of atoms $N_{ex}$ and the magnitude of the relative enhancement. Consider, for example, $p_{z} \propto \exp(-U_{z}/kT)$.
Assume that the asymptotic value of the temperature $T_{0}$ and the optimum (i.e., the minimum)
temperature are different by about two times, i.e,. $T = T_{0}/2$. The asymptotic value of $p_{z}$ is proportional to $\exp(-U_{z}/kT_{0})$
and the value at the optimum temperature is proportional to $\exp(-2U_{0}/kT_{0})$. The ratio of these two $\exp(U_{z}/kT_{0})$
gives the resonant enhancement in the number of atoms in the optimum regime. This enhancement increases with decrease in the temperature $T_{0}$.
%
\section{Chemical potentials}
\begin{figure}
\includegraphics[width=8.6cm]{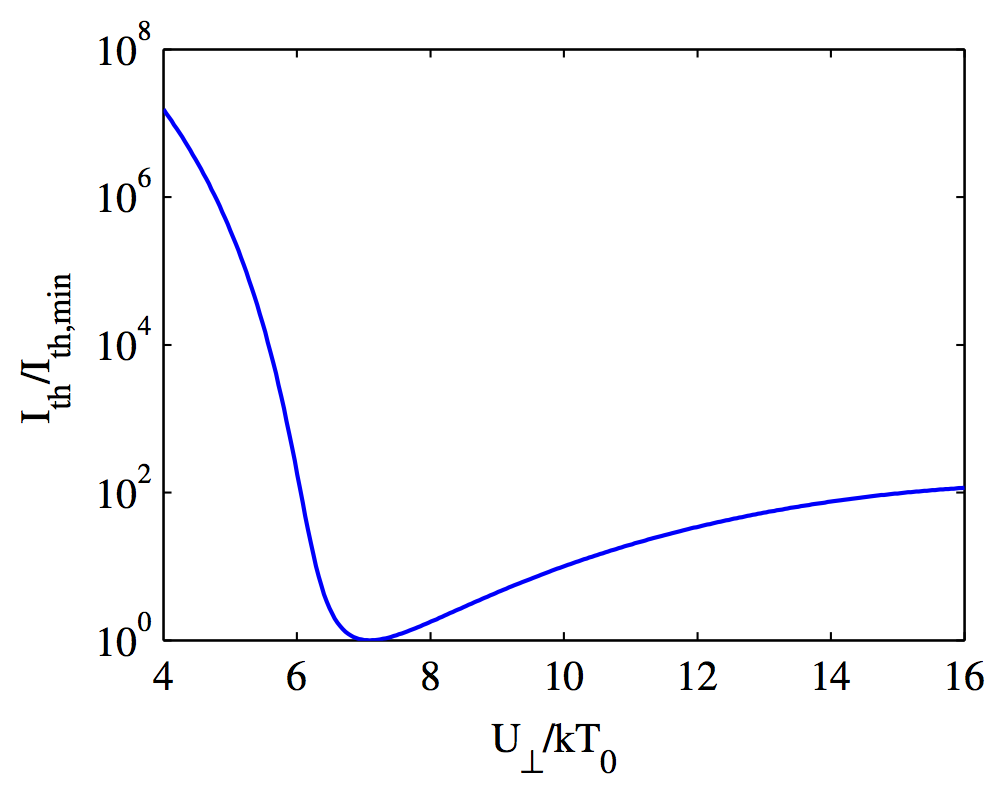}
\caption{\label{fig:I_th_versus_U_p} Normalized threshold incident flux $I_{th}/I_{th,min}$ versus transverse evaporation threshold $U_{\perp}/k T_{0}$}
\end{figure}

%
Above the BEC threshold, the chemical potential of the thermal atoms is described by the expression Eq~(\ref{mu_ex}),
which can be rewritten as
\begin{equation}\label{mu_ex_1}
\frac{\mu_{ex}}{kT_{0}} = \frac{\zeta(3)}{\zeta(2)}\frac{T}{T_{0}}\left[\frac{N_{ex}}{N_{0}}\left(\frac{T_{0}}{T}\right)^{3}\frac{N_{0}}{\zeta(3)}\left(\frac{ \bar{hf}}{kT_{0}}\right)^{3} - 1\right]
\end{equation}
In the presence of BEC, the average collision frequency $\gamma$ Eq.~(\ref{mean_collision_rate_BEC}) is a function of temperature only, and, consequently,
so are the probabilities $p_{z}$ and $p_{\perp}$ given by Eqs.~(\ref{p_z_and_kappa_z}) and (\ref{p_perp_and_kappa_perp}). Thus, the temperature depends on the
parameters of the trap but not on the number of atoms. For fixed parameters of the trap, the number of thermal atoms
$N_{ex}$ is directly proportional to the flux of incident atoms
\begin{equation}
    N_{ex} = \frac{I_{in} - I_{l}}{(p_{\perp} + p_{z})\gamma}.
\end{equation}
allowing us to write Eq.~(\ref{mu_ex_1}) as
\begin{eqnarray}\label{mu_ex_2}
&&\frac{\mu_{ex}}{kT_{0}} =\frac{\zeta(3)}{\zeta(2)}\frac{T}{T_{0}}\nonumber \\
&&\times\left[\frac{I_{in} - I_{l}}{(p_{\perp} + p_{z})\gamma}\left(\frac{T_{0}}{T}\right)^{3}\frac{1}{\zeta(3)}\left(\frac{\bar{hf}}{kT_{0}}\right)^{3} - 1\right]
\end{eqnarray}
It is convenient to introduce the threshold flux for zero load defined by the relation
\begin{eqnarray}\label{I_th}
I_{th} = \zeta(3)\left[p_{\perp}(0) + p_{z}(0)\right]\gamma(0)\left(\frac{kT(0)}{\bar{hf}}\right)^{3},
\end{eqnarray}
where $T(0) = T(I_{l} = 0)$ is the temperature of the thermal atoms at zero load, and so are
$\gamma(0) = \gamma(T(0))$, $p_{\perp}(0)$ and $p_{z}(0)$. The threshold flux $I_{th}$ normalized to its minimum value as a function of the transverse trap height $U_{\perp}$ for all other parameters fixed (and the same as in Figs.~\ref{fig:T_versus_U_p} and \ref{fig:N_versus_U_p}) is shown in Fig.~\ref{fig:I_th_versus_U_p}. The minimum value of the flux corresponds to the same value of $U_{\perp}$ that provides the lowest temperature and the largest number of atoms in Figs.~\ref{fig:T_versus_U_p} and \ref{fig:N_versus_U_p}.

Equation (\ref{mu_ex_2}) in terms of the threshold flux (\ref{I_th}) can be rewritten as
\begin{eqnarray}\label{mu_ex_fin}
&&\frac{\mu_{ex}}{kT(0)}= \frac{\zeta(3)}{\zeta(2)}\frac{T}{T(0)}\nonumber\\
&&\times\left[\frac{I_{in} - I_{l}}{I_{th}} \left(\frac{T(0)}{T}\right)^{3}\frac{[p_{\perp}(0) + p_{z}(0)]\gamma(0)}{(p_{\perp} + p_{z})\gamma}  - 1 \right]\nonumber\\
\end{eqnarray}
Equation (\ref{I_load}) can be rewritten in the form
\begin{equation}\label{I_load_tmp1}
    I_{l} = \frac{(16\pi)^{2}\sqrt{2}ma_{s}{\bar a}}{15 h^{3}}(kT)^{2}\frac{\mu_{ex} - \mu_{a}}{kT}\left(\frac{\mu_{a}}{h\bar{f}}\right)^{5/2}.
\end{equation}
With the help of expression (\ref{mean_collision_rate_BEC}), this equation can be expressed as
\begin{equation}\label{I_load_tmp2}
    \frac{\mu_{ex} - \mu_{a}}{kT}\left(\frac{\mu_{a}}{kT}\right)^{5/2} = \frac{15\zeta(3/2)}{8\sqrt{2}}\frac{a_{s}}{{\bar a}}
    \frac{I_{l}}{\gamma}\left(\frac{h\bar{f}}{kT}\right)^{5/2},
\end{equation}
From Eq.~(\ref{mean_collision_rate_BEC}),
\begin{equation}
    \frac{{\bar f}}{kT} = \frac{a_{s}}{{\bar a}}\left(\frac{{\bar f}}{\gamma}\right)^{1/2}\left[8\zeta(3/2)\right]^{1/2},
\end{equation}
allowing us to rewrite Eq.~(\ref{I_load_tmp2}) in the form
\begin{eqnarray}\label{equation_for_mu_a}
    \frac{\mu_{ex} - \mu_{a}}{kT}\left(\frac{\mu_{a}}{kT}\right)^{5/2} =  \frac{15\zeta(3)}{16}\left(\frac{\zeta^{3}(3/2)}{4\pi}\right)^{1/4} \nonumber \\
    \left(\frac{a_{s}}{{\bar a}}\right)^{1/2}
    \left(\frac{\gamma}{{\bar f}}\right)^{1/2}[p_{\perp}(0) + p_{z}(0)]\frac{I_{l}}{I_{th}}\left(\frac{T(0)}{T}\right)^{5}
\end{eqnarray}
The numerical coefficient $(15\zeta(3)/16)[\pi \zeta^{3}(3/2)]^{1/4} \approx 1.2.$

Finally, the number of the BEC atoms follows from Eq.~(\ref{mu_a}):
\begin{equation}
    N_{a} = \frac{\bar{a}}{a_{s}}\frac{2^{5/2}}{15}\left(\frac{\mu_{a}}{kT}\right)^{5/2}\left(\frac{kT}{\bar{hf}}\right)^{5/2}.
\end{equation}
%

\section{Influence of the load}
\begin{figure}
\includegraphics[width=8.6cm]{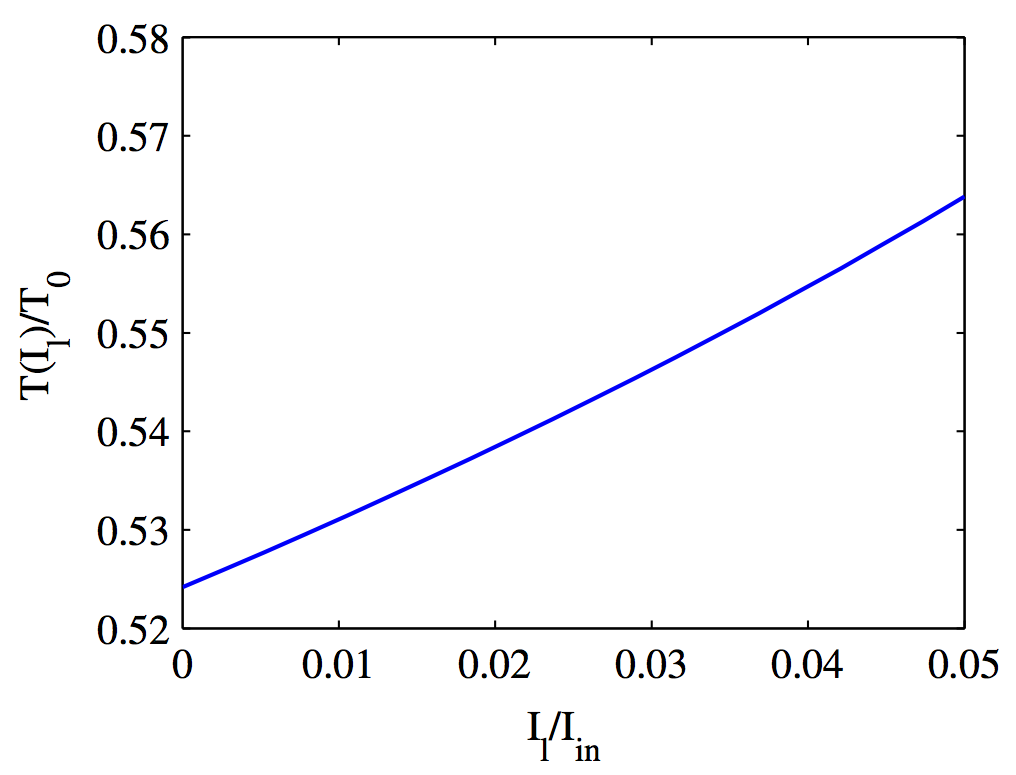}
\caption{\label{fig:T_versus_load_same_U_p} Normalized minimum temperature $T(I_{l})/T_{0}$ versus load $I_{l}/I_{in}$ for fixed value of $U_{\perp}$.}
\end{figure}

\begin{figure}
\includegraphics[width=8.6cm]{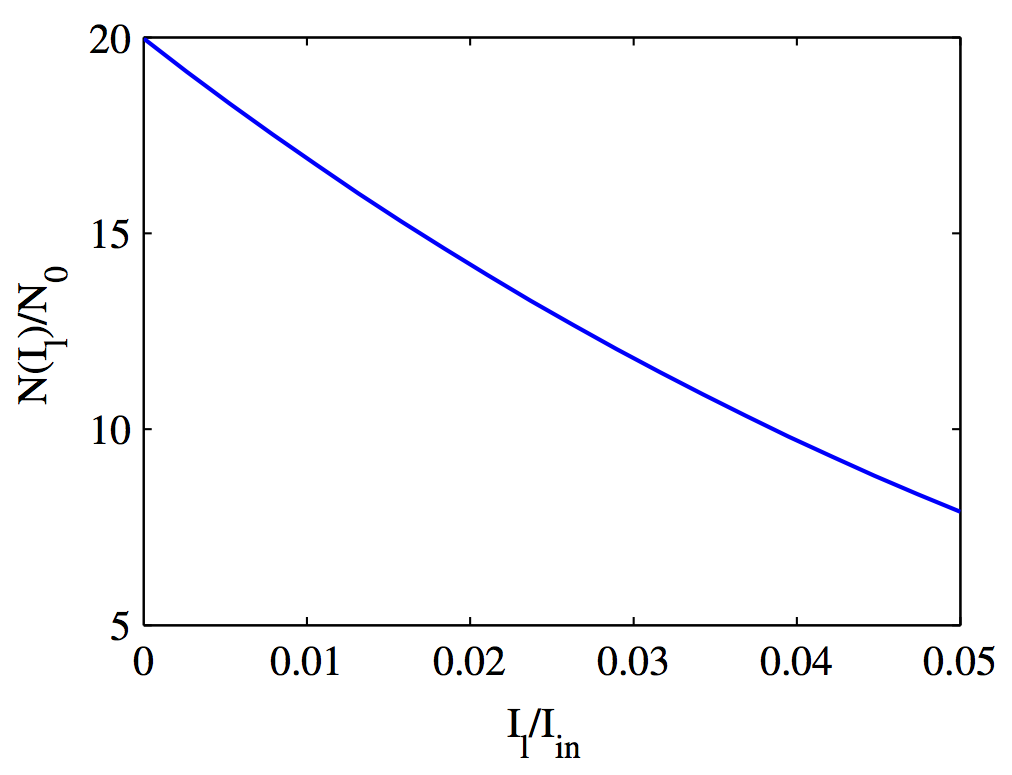}
\caption{\label{fig:N_versus_load_same_U_p} Normalized maximum number of thermal atoms $N(I_{l})/N_{0}$ versus load $I_{l}/I_{in}$ for fixed value of $U_{\perp}$.}
\end{figure}

%
Nonzero load $I_{l}$ decreases the maximum value of the thermal atoms and increases the minimum temperature that can be achieved
for a given level of external pumping $I_{in}$ as illustrated by Figs.~\ref{fig:T_versus_load_same_U_p} and \ref{fig:N_versus_load_same_U_p}. Figure \ref{fig:T_versus_load_same_U_p} shows the minimum value of the temperature $T(I_{l})$ that can be achieved for a nonzero load normalized to its asymptotic value $T_{0}$ versus the normalized load $I_{l}/I_{in}$. Figure~\ref{fig:N_versus_load_same_U_p} shows the maximum number of the thermal atoms $N(I_{l})$
normalized to $N_{0}$ versus the normalized load $I_{l}/I_{in}$. For both graphs the values of $U_{z}$, $U_{\perp}$ and $\epsilon$ are kept constant. The value of $U_{\perp}$ corresponds to the optimum value in case of zero load $I_{l} = 0$. i.e., to the value giving the minimum temperature and the maximum number of atoms in Figs.~\ref{fig:T_versus_U_p} and \ref{fig:N_versus_U_p}.
\begin{figure}
\includegraphics[width=8.6cm]{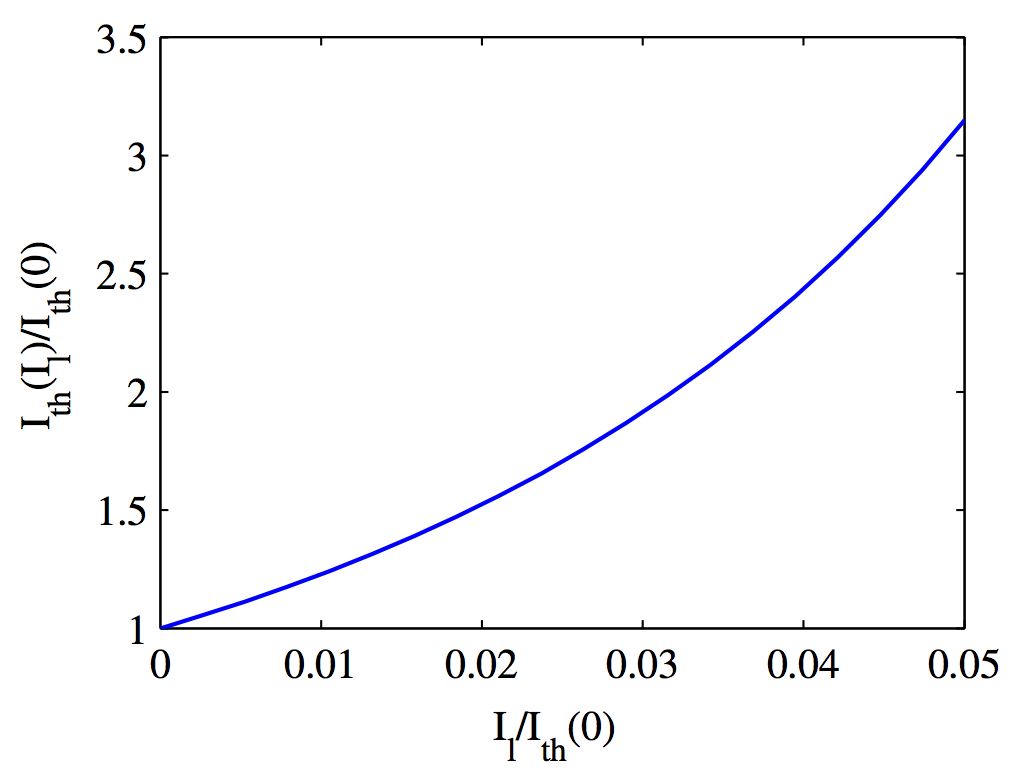}
\caption{\label{fig:I_th_versus_load_same_U_p} Threshold incident flux $I_{th}(I_{l})/I_{th}(0)$ versus normalized load $I_{l}/I_{th}(0)$.}
\end{figure}
\begin{figure}
\includegraphics[width=8.6cm]{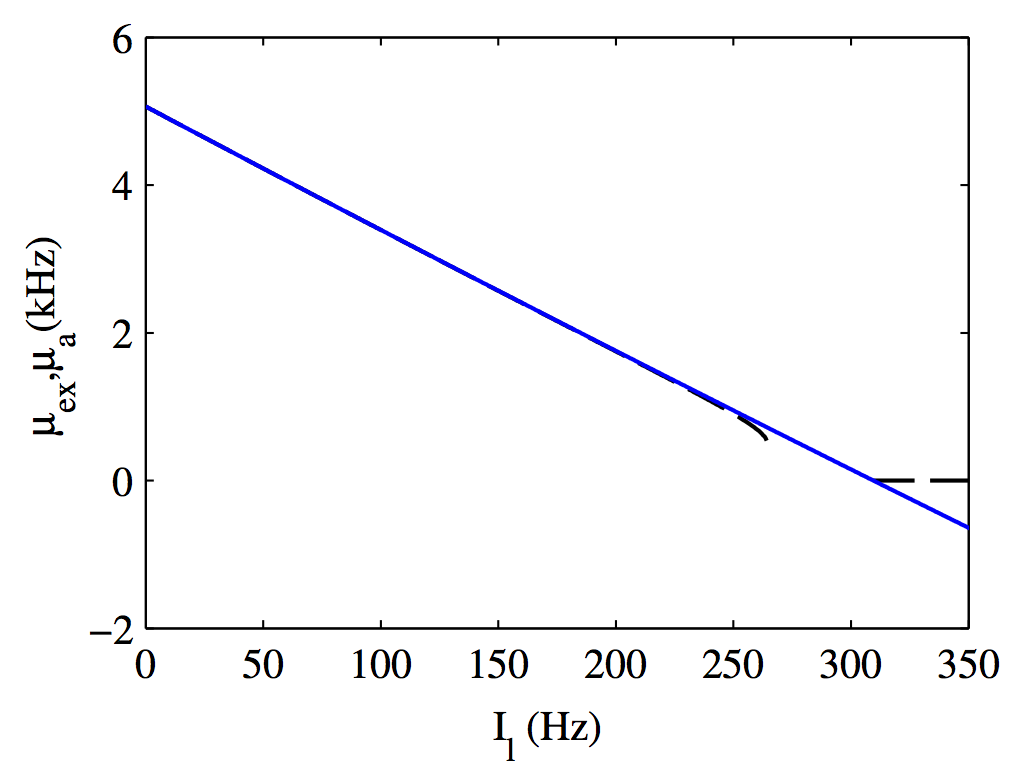}
\caption{\label{fig:mus_abs_versus_load_same_U_p} Chemical potential of thermal atoms $\mu_{ex}$ (solid) and the condensate $\mu_{a}$ (dashed) versus load $I_{l}$ for $I_{in}/I_{th} = 1.1$.}
\end{figure}
Increase in the temperature and decrease in the number of atoms translate in an increase in the threshold flux $I_{th}(I_{l})$ as compared to its value for zero load $I_{th}(0)$ for all other parameters fixed. This is illustrated by Fig.~\ref{fig:I_th_versus_load_same_U_p}.

Increase in the temperature and decrease in the number of atoms seen in (\ref{fig:T_versus_load_same_U_p}) and (\ref{fig:N_versus_load_same_U_p}) are partially due to the fact that the optimum value of $U_{\perp}$ depends on the load. To investigate this circumstance we have also carried out calculations where, for each given value of the load $I_{l}$,  the values of $T$ and $N$ have been optimized with respect to the value of $U_{\perp}$. In other words, $T(I_{l})$ and $N(I_{l})$ were chosen to correspond to the minimum of the temperature and the maximum of the number of atoms on graphs analogous to Figs.~\ref{fig:T_versus_U_p} and \ref{fig:N_versus_U_p}.  The results were very similar to those shown in Figs.~\ref{fig:T_versus_load_same_U_p} and \ref{fig:N_versus_load_same_U_p},  but changes were somewhat less pronounced.

Above the condensation threshold, increasing the load causes increase in the temperature and decrease in the number of thermal atoms.
At some maximum value of the load the system goes below the threshold.
Figure \ref{fig:mus_abs_versus_load_same_U_p} shows the chemical potentials $\mu_{ex}$ and $\mu_{a}$ versus the load $I_{l}$. The calculations have been carried out for the trap with $f_{z} = 100~\mathrm{Hz}$, $f_{\perp} = 2~\mathrm{kHz}$, $\epsilon = 0.7$ and for the zero-load input flux $I_{in} \approx 6.7 \times 10^{4}~\text{atoms/s}$. This value of the flux is ten percent above the threshold of formation of the BEC, i.e., $I_{in}/I_{th} = 1.1$ and corresponds to the threshold number of thermal atoms for zero load $N_{ex} \approx 10^{6}$. The results are presented in frequency units obtained by dividing the chemical potentials by $h$, where $h$ is Planck's constant. The chemical potential of the condensate lies slightly below that of the thermal atoms and closely follows it for most values of the load.
The difference between $\mu_{a}$ and $\mu_{th}$ increases with larger loads. In a narrow region around the maximum load above the condensation threshold stationary solutions of Eqs.~(\ref{steady_state_N}), (\ref{steady_state_T}) do not exist as is indicated by a gap in the graph of $\mu_{a}$. The stationary solutions disappear when the difference $\mu_{ex} - \mu_{a}$ reaches $(2/7)\mu_{ex}.$ While the presence of the gap is interesting, we do not address it further here.

%
\section{Equivalent Circuits}
Having established the physics of its behavior we are now in a position to construct an equivalent circuit for the battery. Given a fixed trap potential there are three adjustable circuit operating parameters: the input flux $I_{in}$, the excess energy $\epsilon$ (or input power), and the load current $I_{l}$. It  is natural to associate the input atom flux with a current source, and from there the behavior of the system can be cast in terms of the five element circuit shown in Fig.~(\ref{fig:circuit_model}).  While our analysis is valid under certain assumptions such as small normalized chemical potential, the equivalent circuit more generally describes the small-signal behavior of the battery about any operating point $\{\tilde{I}_{in},\tilde{I}_{l},\tilde{\mu}_{ex}\}$. The shunt resistance $R_p$ is determined by the dependence of the thermal chemical potential on the input current:
\begin{figure}
\includegraphics[width=8.6cm]{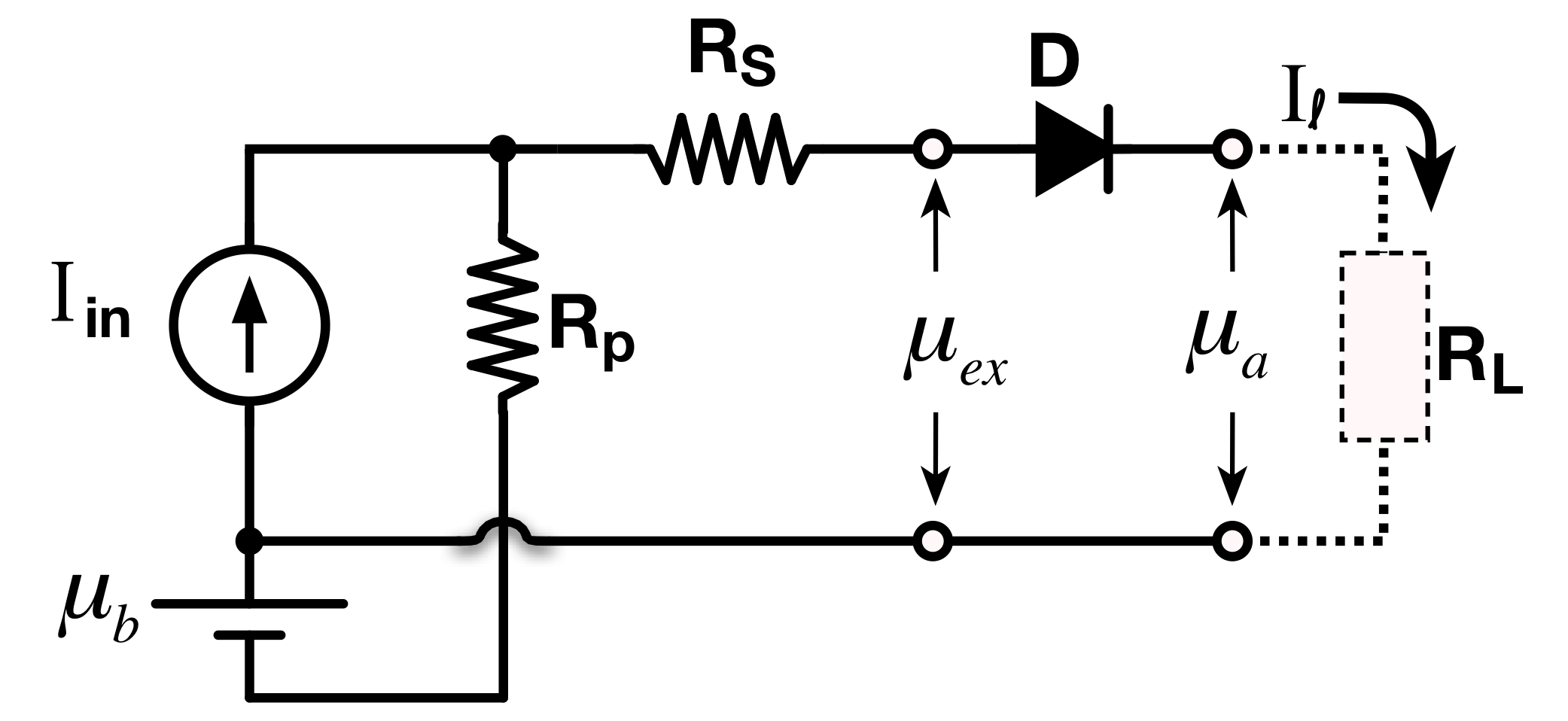}
\caption{\label{fig:circuit_model} Equivalent circuit model of the atomtronic battery.}
\end{figure}
\begin{equation}\label{R_p}
    R_{p}= \left.{\frac{\partial\mu_{ex}}{{\partial}{I_{in}}}}\right|_{\{\tilde{I}_{in},\tilde{I}_{l},\tilde{\mu}_{ex}\}},
\end{equation}
while the series resistance is determined by the dependence of the thermal chemical potential on the load current:
\begin{equation}\label{R_s}
    R_{s}= -  \left.{\left({\frac{\partial\mu_{ex}}{{\partial}{I_{l}}}+R_{p}}\right)}\right|_{\{\tilde{I}_{in},\tilde{I}_{l},\tilde{\mu}_{ex}\}}.
\end{equation}
The bias chemical potential is determined from:
\begin{equation}\label{mu_b}
    \mu_{b}=\tilde{I}_{in}R_{p}-\tilde{I}_{l}R_s-\tilde{\mu}_{ex}.
\end{equation}
Near the threshold input current the bias can be determined simply by $\mu_{b}=I_{th}R_{p}$.

\begin{figure}
\includegraphics[width=8.6cm]{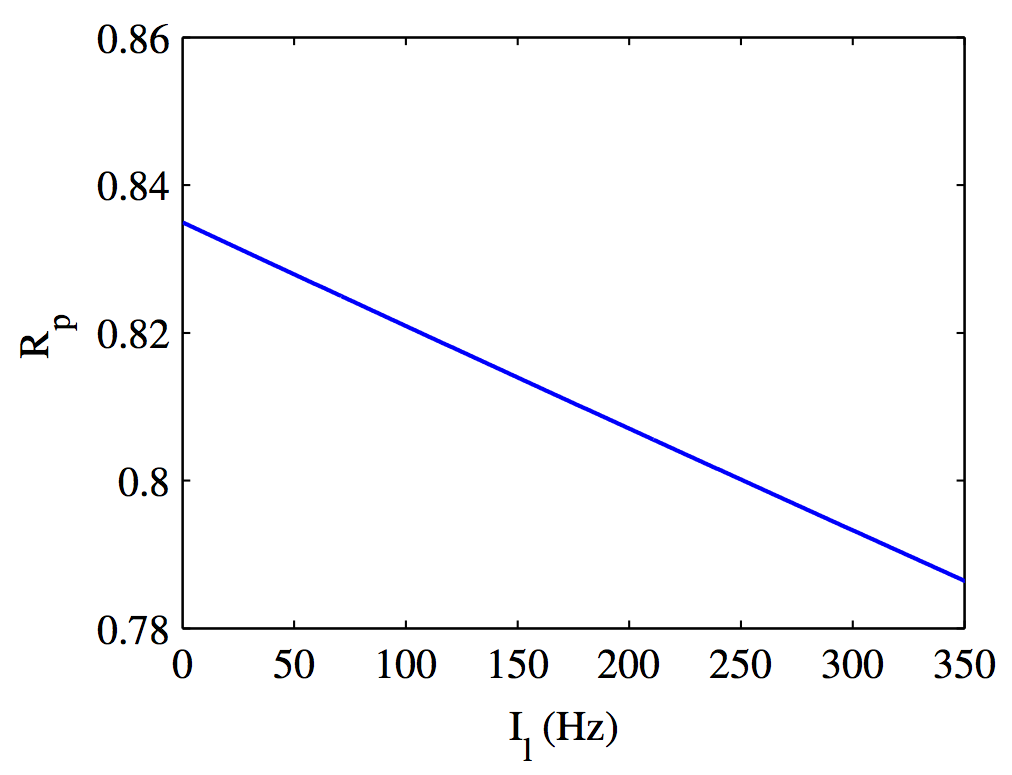}
\caption{\label{fig:Rp_versus_load} Resistance $R_{p}$ versus load $I_{l}$ for $I_{in}/I_{th} = 1.1$.}
\end{figure}

\begin{figure}
\includegraphics[width=8.6cm]{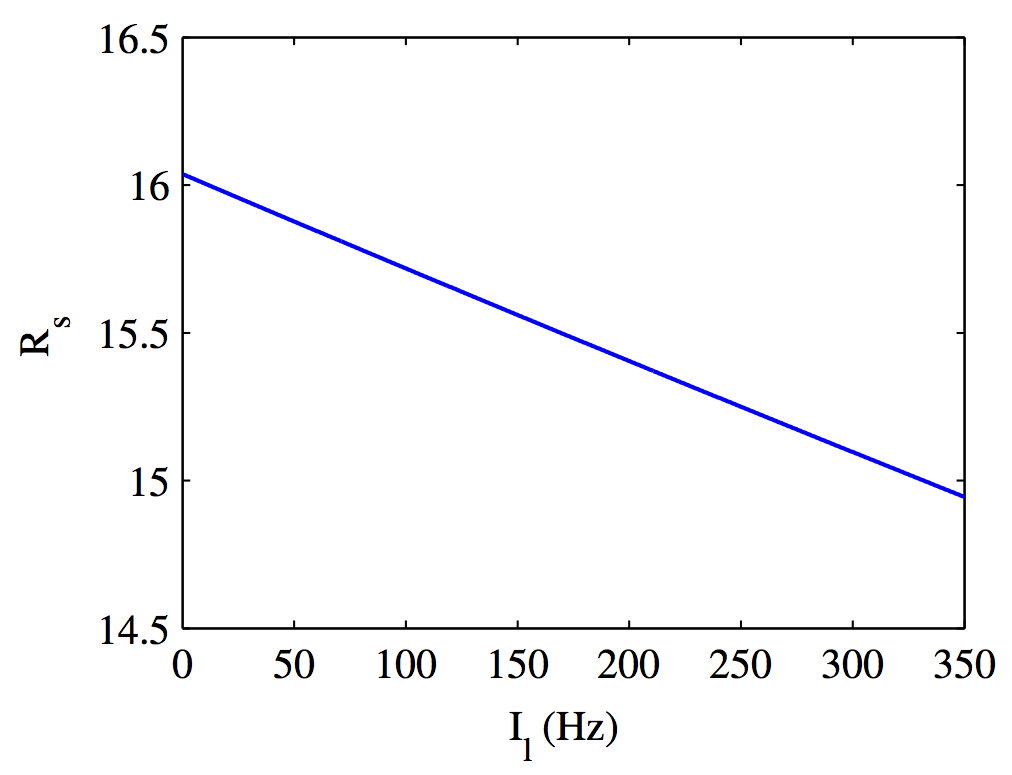}
\caption{\label{fig:Rs_versus_load} Resistance $R_{s}$ versus load $I_{l}$ for $I_{in}/I_{th} = 1.1$.}
\end{figure}

\begin{figure}
\includegraphics[width=8.6cm]{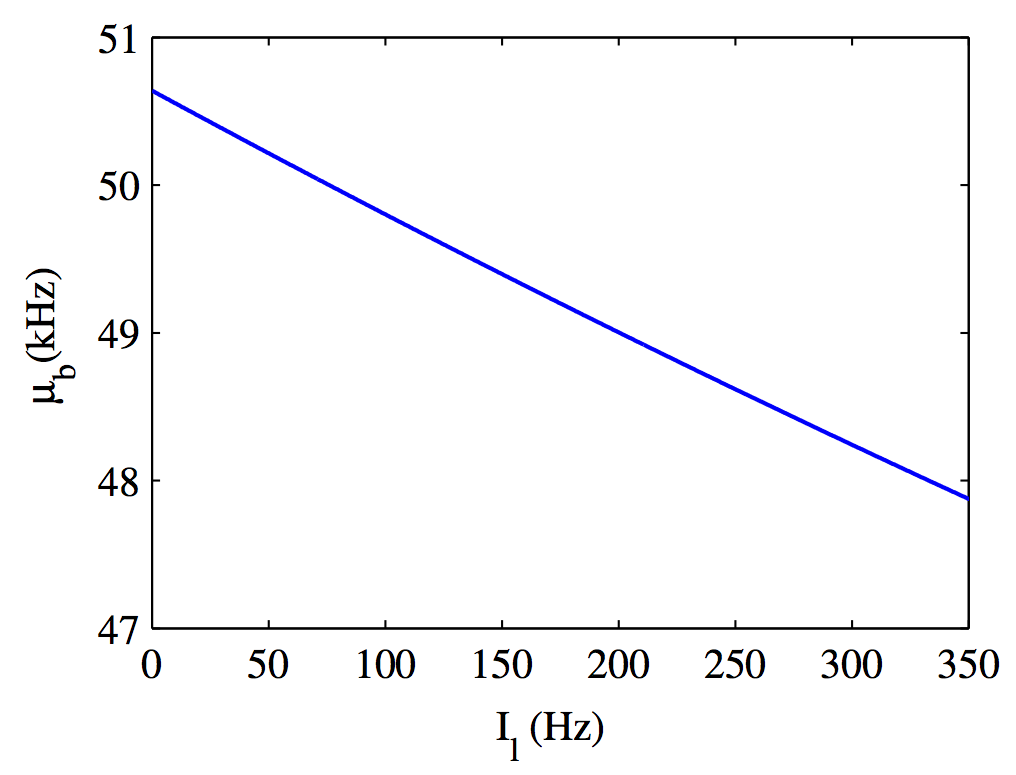}
\caption{\label{fig:mu_b_abs_versus_load} Bias potential $\mu_{b}$ versus load $I_{l}$ for $I_{in}/I_{th} = 1.1$.}
\end{figure}

\begin{figure}
\includegraphics[width=8.6cm]{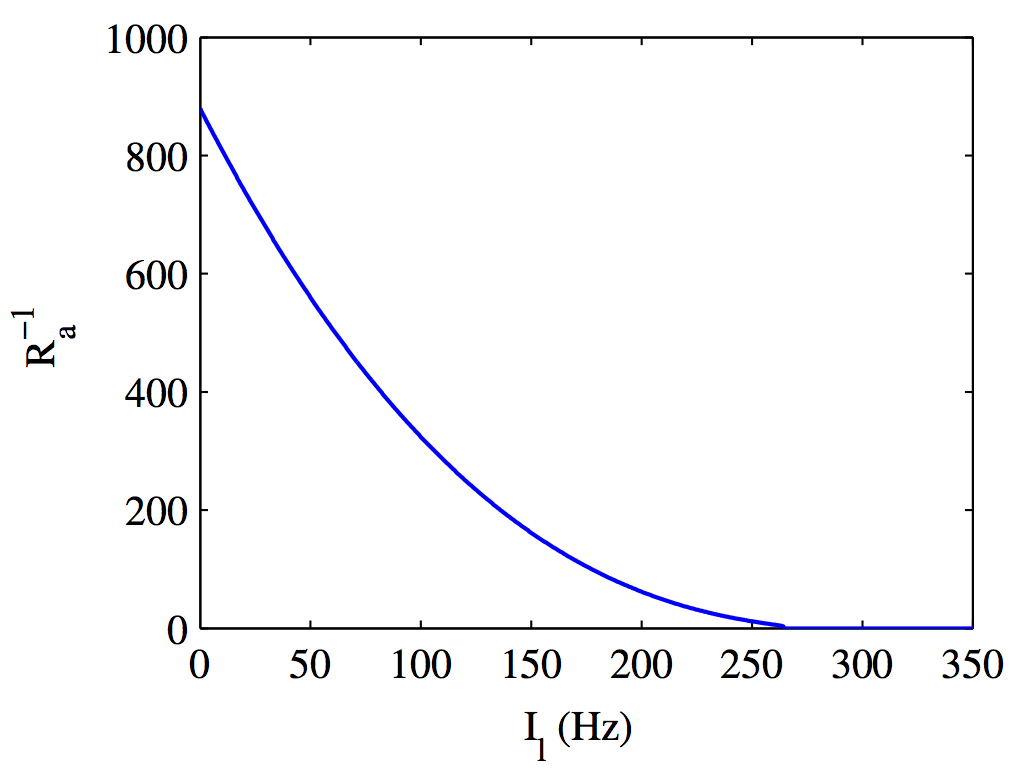}
\caption{\label{fig:inverse_Ra} Inverse of the resistance $R_{a}$ versus load $I_{l}$ for $I_{in}/I_{th} = 1.1$.}
\end{figure}

The series resistance $R_{s}$ should nominally be followed by the resistance $R_{a}$ defined by the relation (\ref{R_a}) to account for the drop between the chemical potential of the thermal atoms and that of the condensate.  Fig.~\ref{fig:inverse_Ra} shows the graph of $R_{a}^{-1}$ versus the normalized load for the same parameters as those of Fig.~(\ref{fig:mus_abs_versus_load_same_U_p}). At low values of load current the conductance is large and the resistance is small compared with $R_{s}$.  As the load current approaches the point for which the condensate ceases to exist, the resistance increases rapidly.  In the circuit model we have chosen to represent the role of the resistance $R_{a}$ with a diode.  At finite temperature a diode, like the resistance $R_{a}$, presents a conductivity that is a strong function of the forward-biased potential across it; for reverse bias it presents zero conductivity,

Figs.~\ref{fig:Rp_versus_load}, \ref{fig:Rs_versus_load}, and \ref{fig:mu_b_abs_versus_load} show the graphs of $R_{p}$, $R_{s}$ and $\mu_{b}$ defined by the relations (\ref{R_p})-(\ref{mu_b}), respectively, versus the load $I_{l}$.  All parameters are the same as for Fig.~\ref{fig:mus_abs_versus_load_same_U_p}. With the chemical potentials reported in frequency units resistance is dimensionless. The circuit parameters vary about $6\%$, relatively little over the span of output currents, indicating that the circuit model accurately reflects the analytical model to this level.    The chemical potential $\mu_{ex}$ becomes zero for $I_{l} = 310$ Hz.
%

\begin{figure}
\includegraphics[width=8.6cm]{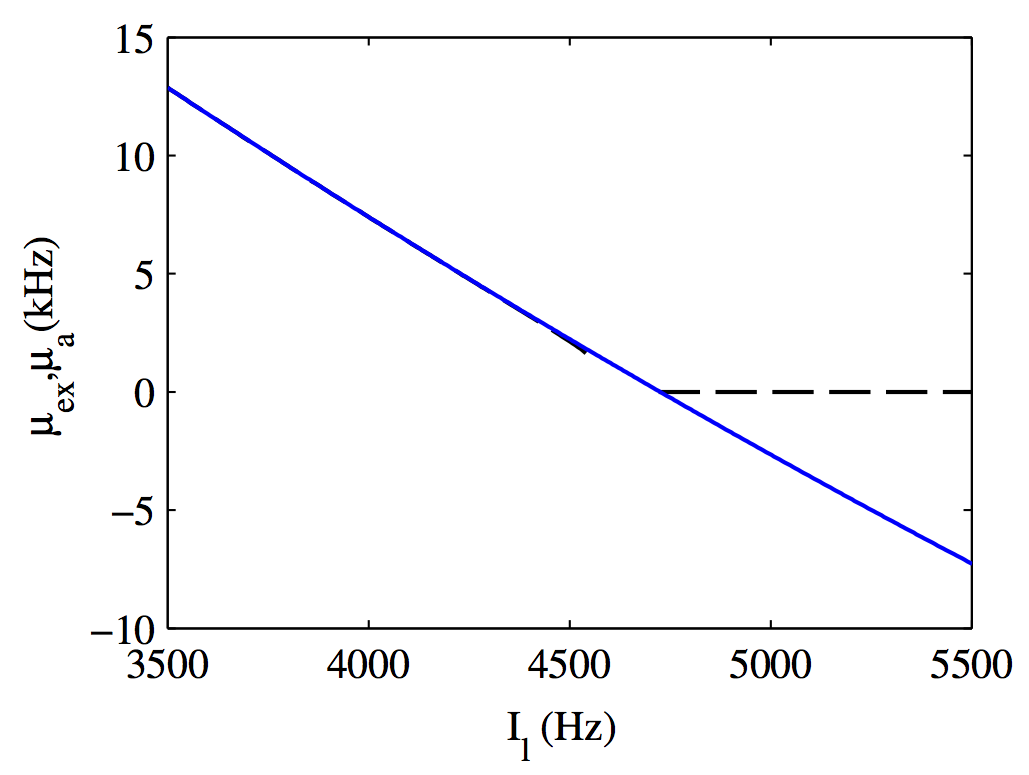}
\caption{\label{fig:mus_abs_versus_load_large} Chemical potential of thermal atoms $\mu_{ex}$ (solid) and the condensate $\mu_{a}$ (dashed) versus load $I_{l}$ for $I_{in}/I_{th} = 2.2$.}
\end{figure}

\begin{figure}
\includegraphics[width=8.6cm]{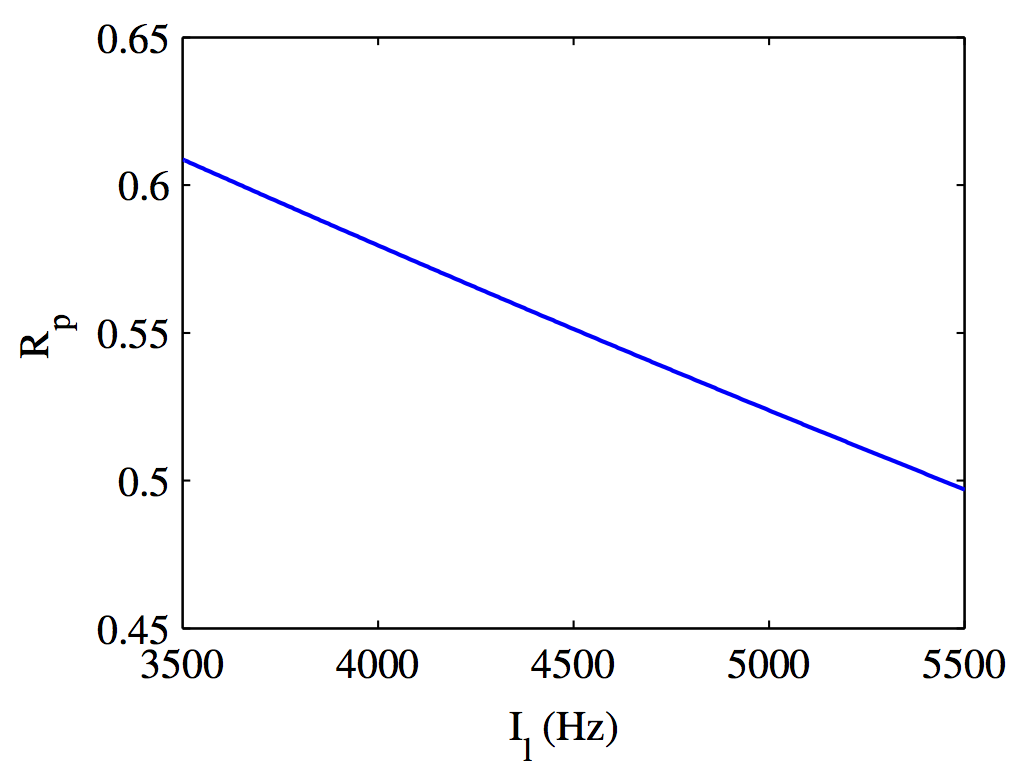}
\caption{\label{fig:Rp_versus_load_large} Resistance $R_{p}$ versus load $I_{l}$ for $I_{in}/I_{th} = 2.2$.}
\end{figure}

\begin{figure}
\includegraphics[width=8.6cm]{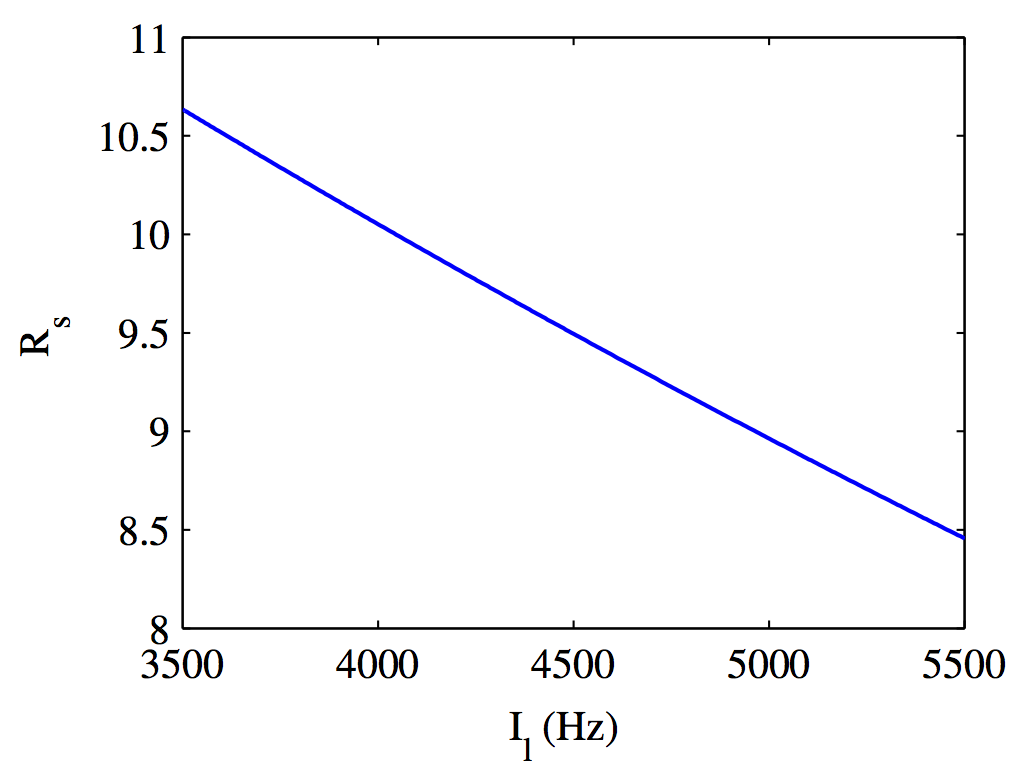}
\caption{\label{fig:Rs_versus_load_large} Resistance $R_{s}$ versus load for $I_{l}$ for $I_{in}/I_{th} = 2.2$.}
\end{figure}

\begin{figure}
\includegraphics[width=8.6cm]{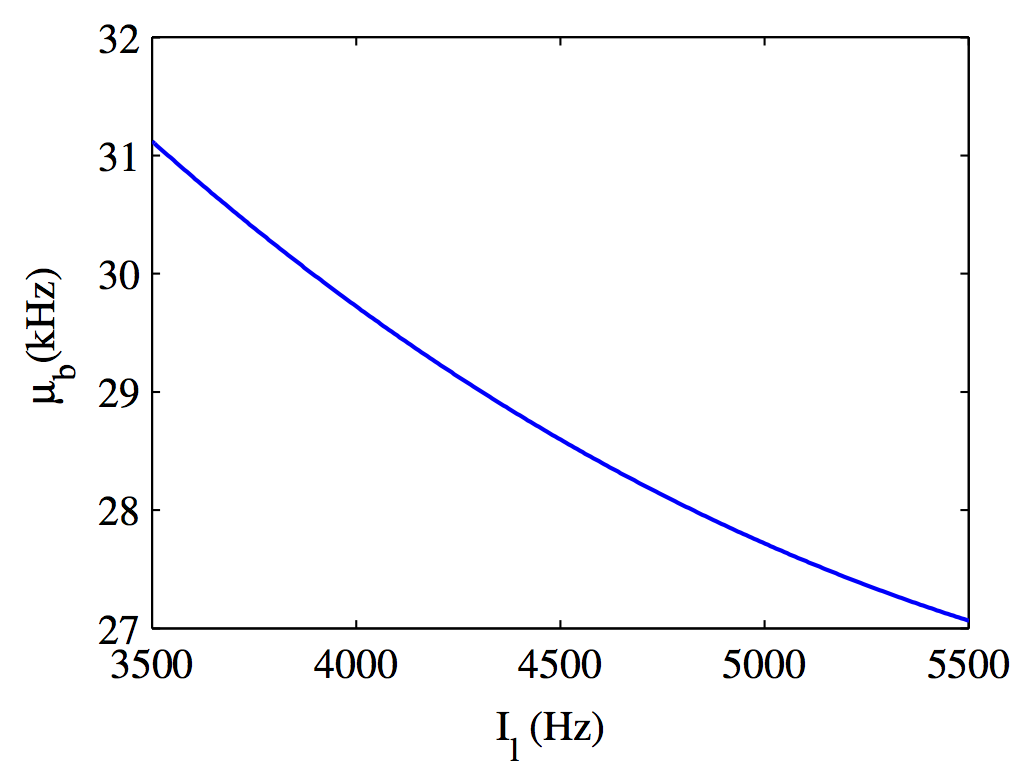}
\caption{\label{fig:mu_b_abs_versus_load_large} Bias potential $\mu_{b}$ versus load $I_{l}$ for $I_{in}/I_{th} = 2.2$.}
\end{figure}

Figs.~\ref{fig:mus_abs_versus_load_large}, \ref{fig:Rp_versus_load_large}, \ref{fig:Rs_versus_load_large}, and \ref{fig:mu_b_abs_versus_load_large} show the graphs of $\mu_{ex}$, $\mu_{a}$, $R_{p}$, $R_{s}$ and $\mu_{b}$, versus the load $I_{l}$. These graphs present the same quantities as Figs.~\ref{fig:mus_abs_versus_load_same_U_p}, \ref{fig:Rp_versus_load}, \ref{fig:Rs_versus_load}, and \ref{fig:mu_b_abs_versus_load}, but for the value of the incident flux that is twice larger, i.e., $I_{in}/I_{th} = 2.2$. Since our model is valid in the limit $\mu_{ex}, \mu_{a} \ll kT$, only the region of large loads when the system is not too far away from the threshold is shown.  In this case the circuit element values vary as much as about $20\%$ over the range of output currents. The resistance $R_{a}$ will have a still smaller impact on the effective series resistance.  We note that the chemical potential at zero load significantly increases and the series resistance significantly decreases with twice the input power.  As discussed below, this means that both the maximum output power and the efficiency of the battery improve with increased input flux.

\begin{figure}
\includegraphics[width=8.6cm]{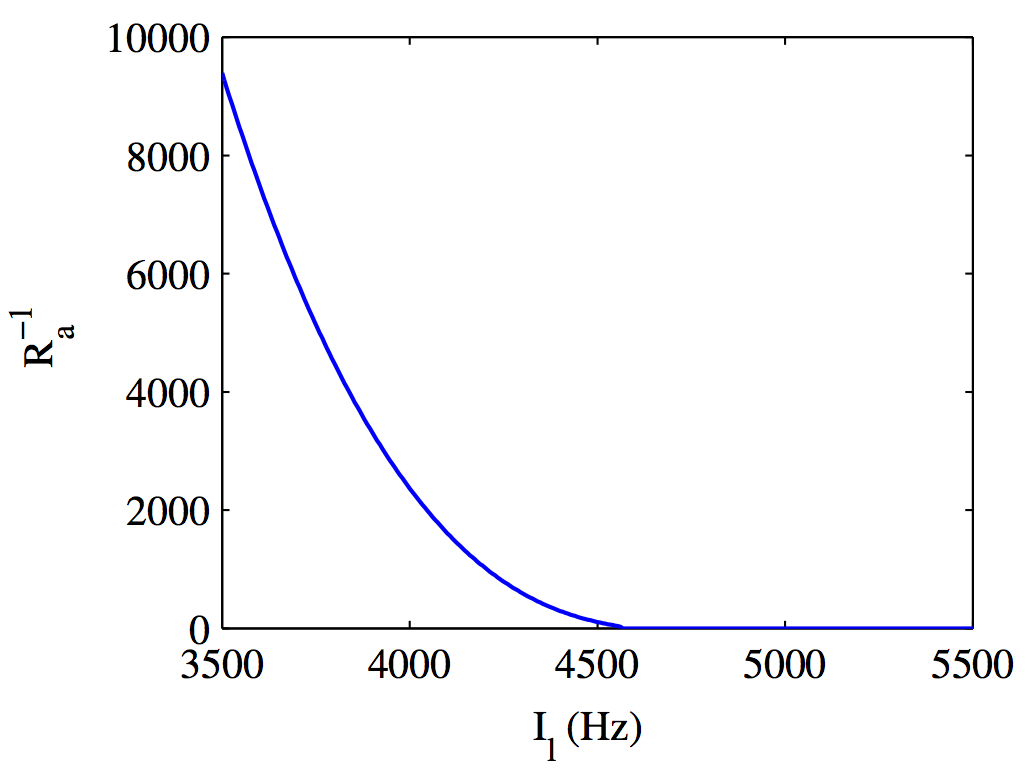}
\caption{\label{fig:inverse_Ra_large} Inverse of the resistance $R_{a}$ versus load $I_{l}$ for $I_{in}/I_{th} = 2.2$.}
\end{figure}
\begin{figure}
\includegraphics[width=5.6cm]{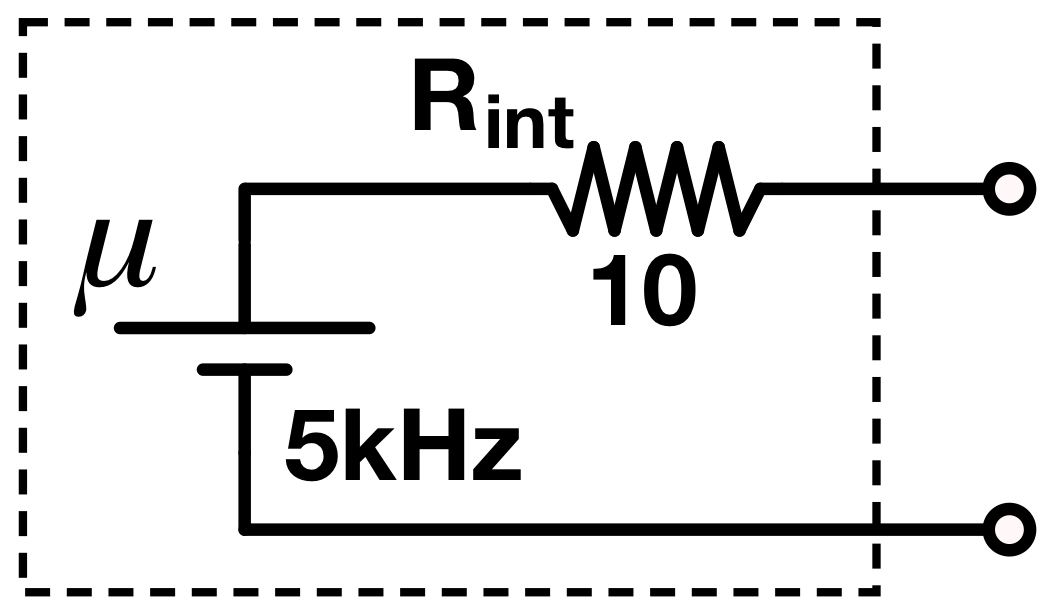}
\caption{\label{fig:Voltage_Source} Th\'{e}venin equivalent voltage source with typical element values based on this work.}
\end{figure}

A Th\'{e}venin equivalent circuit hides the complexity and extracts the essence of the battery as a self-contained source of power.  Fig.~\ref{fig:Voltage_Source} shows a Th\'{e}venin equivalent voltage source with values typical of this work. Referencing the equivalent circuit Fig.~\ref{fig:circuit_model} and Eqns.~(\ref{R_p})--(\ref{mu_b}) the potential is given by $\mu=\tilde{I}_{in}R_{p}-\mu_{b}$ and the internal resistance $R_{int}=R_{s}$. The simple diagram elegantly encapsulates the performance one can expect from the battery, namely, that it will supply a maximum power of $P_{max}=(h\mu)^{2}/4R_{int}$ to the load through a flux of atoms $I_{l}=\mu/2R_{int}=250$Hz having energy $E=h\mu/2$ per atom.  As it delivers the maximum power it will produce heat at a rate \textit{at least} equal to the maximum power.  At finite temperature, the internal resistance will also impose noise on the output flux, although a discussion of this atomtronic equivalent of Johnson noise is beyond what we address here.
%
\section{Conclusions}
This work has considered the behavior of a highly asymmetric trap as a means of supplying power to an atomtronic circuit.  The trap potential is characterized by four parameters: its longitudinal and transverse frequencies $f_{z}$, $f_{\perp}$, and the corresponding trap heights, $U_{z}$, $U_{\perp}$.  The battery operating point is set by the input flux $I_{in}$ and power, $P_{in}=I_{in}(1+\epsilon)U_{\perp}$ as well as the load flux $I_{l}$.  Among the set points the battery performance is most strongly affected by the excess energy $\epsilon$ which exhibits a resonant optimum for a given set of trap parameters.

We have shown that the battery can be modeled in terms of an equivalent circuit.  In particular for a specific operating point the battery can  be modeled with its Th\`{e}venin equivalent chemical potential and finite internal resistance. As is true of its electrical counterpart the battery is ideally designed with its intended load in mind, i.e., that the trap parameters are chosen such that power is optimally transferred from the input flux to the output flux.    Our analysis reveals that the ratio of input to output flux is in the range of a few hundred to one and generally improves with larger excess energy $\epsilon$.  One is often more interested in the power efficiency: the validity of our analysis is restricted to parameter regimes outlined in Section (\ref{thermal_cloud}) and therefore so are there limitations on the prediction of power optimization. In the context of the Th\`{e}venin equivalent circuit, maximum power efficiency is obtained by matching the internal to the load resistance.

The ability of the battery to supply condensed atoms to a load has been analyzed here without specifying how the load is implemented: it is effectively treated in terms of a lumped resistance.  It is worth keeping in mind that in general a load will present a complex impedance to the battery, and that the reactive components will impact battery behavior.  Along the same lines, although we depict circuits with lumped elements, the deBroglie wavelength of the atoms is on the order or smaller than the physical dimensions of the circuits.  Circuits are therefore more aptly described in terms of transmission lines and waveguides, linear and nonlinear, than they are in terms of lumped elements. Atomtronic circuitry is thus more akin to the microwave domain of electronics than it is to the audio domain.

We have focused on a particular means of implementing a battery for atomtronics, yet it will be true that \emph{any} means of supplying power to a circuit can be modeled in terms of a Th\'{e}venin equivalent source, at least over some small signal regime around a quiescent point.  The significant conclusion, then, is that any power source will internally dissipate heat in its internal resistance, deliver a certain maximum amount of power to a load, and impose noise onto the circuit.  Equivalent circuits place a possibly diverse set of battery implementations on an equal footing in terms of their ability to drive a circuit.  In general it may be difficult to predict the parameters of an equivalent circuit, but it should be possible to measure them for any specific realization of a battery.


\section{Acknowledgements}
This work was supported by the Office of Scientific Research, by the National Science Foundation and by the Charles Stark Draper Laboratories.

\bibliography{Battery}

\newpage

\end{document}